\documentclass[runningheads]{llncs}
\usepackage[T1]{fontenc}
\usepackage{multirow}

\usepackage{pgfgantt}
\usepackage{lscape}
\usepackage{graphicx}
\usepackage{comment}
\usepackage{booktabs}
\usepackage{amsmath}
\usepackage{pgfplots}
\usepackage{tikz}
\usetikzlibrary{fit,calc,shapes,arrows,matrix,arrows.meta}
\usepackage{booktabs}
\usepackage{tabularx}
\usepackage{hyperref}
\usepackage{subcaption}
\usepackage{xargs}                      
\usepackage{xcolor}
\usepackage[colorinlistoftodos,prependcaption,textsize=small]{todonotes}
\newcommandx{\unsure}[2][1=]{\todo[linecolor=red,backgroundcolor=red!25,bordercolor=red,#1]{#2}}
\newcommandx{\change}[2][1=]{\todo[linecolor=blue,backgroundcolor=blue!25,bordercolor=blue,#1]{#2}}
\newcommandx{\info}[2][1=]{\todo[linecolor=OliveGreen,backgroundcolor=OliveGreen!25,bordercolor=OliveGreen,#1]{#2}}
\newcommandx{\improvement}[2][1=]{\todo[linecolor=magenta,backgroundcolor=magenta!25,bordercolor=magenta,#1]{#2}}
\newcommandx{\mytodo}[2][1=]{\todo[linecolor=red,backgroundcolor=red!25,bordercolor=red,#1]{TODO: #2}}
\newcommandx{\apurvtodo}[2][1=]{\todo[linecolor=red,backgroundcolor=red!25,bordercolor=red,#1]{Apurv's TODO: #2}}
\newcommandx{\thiswillnotshow}[2][1=]{\todo[disable,#1]{#2}}

\definecolor{C0}{HTML}{1f77b4}
\definecolor{C1}{HTML}{ff7f0e}
\definecolor{C2}{HTML}{2ca02c}
\definecolor{C3}{HTML}{d62728}
\definecolor{C4}{HTML}{9467bd}
\definecolor{C5}{HTML}{8c564b}
\definecolor{C6}{HTML}{e377c2}
\definecolor{C7}{HTML}{7f7f7f} 
\definecolor{C8}{HTML}{bcbd22}
\definecolor{C9}{HTML}{17becf}
\begin{document}
\title{SProBench: Stream Processing Benchmark\\ for High Performance Computing Infrastructure}
\titlerunning{SProBench}
%
\author{%
    Apurv Deepak Kulkarni \and Siavash Ghiasvand
}
\authorrunning{
    A. Kulkarni and S. Ghiasvand
}
%
\institute{
    Center for Scalable Data Analytics and Artificial Intelligence (ScaDS.AI)\\
    Center for Interdisciplinary Digital Sciences (CIDS) \\
    TUD Dresden University of Technology, Dresden, Germany \\
    \email{\{apurv.kulkarni,siavash.ghiasvand\}@tu-dresden.de}
}
\maketitle              
\begin{abstract}
Recent advancements in data stream processing frameworks have improved real-time data handling, however, scalability remains a significant challenge affecting throughput and latency.
While studies have explored this issue on local machines and cloud clusters, research on modern high-performance computing (HPC) infrastructures is yet limited due to the lack of scalable measurement tools.
This work presents SProBench, a novel benchmark suite designed to evaluate the performance of data stream processing frameworks in large-scale computing systems.
Building on best practices, SProBench incorporates a modular architecture, offers native support for SLURM-based clusters, and seamlessly integrates with popular stream processing frameworks such as Apache Flink, Apache Spark Streaming, and Apache Kafka Streams.
Experiments conducted on HPC clusters demonstrate its exceptional scalability, delivering throughput that surpasses existing benchmarks by more than tenfold.
The distinctive features of SProBench, including complete customization options, built-in automated experiment management tools, seamless interoperability, and an open-source license, distinguish it as an innovative benchmark suite tailored to meet the needs of modern data stream processing frameworks.

\end{abstract}

%
%
\section{Introduction}

Modern big data processing is built around two main paradigms: batch processing and stream processing.
Batch processing involves collecting, storing, and processing data in scheduled intervals, while stream processing processes data as it arrives, enabling real-time analysis.
This real-time capability supports timely decision-making and ensures seamless operations.
With data being generated, processed, and shared at unprecedented rates, often exceeding gigabytes per second~\cite{explodingtopicsAmountData}, raw data typically undergoes significant processing and filtering to reduce its size, making analysis more efficient and manageable.
Data Stream Processing (DSP) focuses on processing data immediately upon acquisition, eliminating the need for intermediate storage.

In recent years, DSP frameworks have advanced significantly, mostly driven by hardware innovations.
Frameworks like Apache Flink~\cite{Katsifodimos2016}, Apache Spark Streaming~\cite{apacheApacheSparkx2122}, and Apache Kafka Stream~\cite{apacheApacheKafka} have improved in areas such as memory management~\cite{databricksWhatSparkTungsten,apacheFlinkMemoryManagement}, support for diverse databases, edge computing compatibility~\cite{Volker2024}, and efficient state management.
Each framework specializes in specific aspects of stream processing.
Given the diversity of DSP tasks, the setup of these frameworks and the underlying hardware are critical for ensuring the functionality of the entire stream processing pipeline.
Benchmarking these frameworks against modern stream processing demands provides valuable insights into their performance across various metrics.
It highlights strengths, identifies limitations and bottlenecks, and helps select the most suitable framework for specific tasks.
Additionally, benchmarking reveals how effectively the underlying hardware is utilized.

Although stream processing frameworks are increasingly relied upon, there are notable gaps in existing comparative studies of these frameworks.
One major limitation is the lack of evaluations conducted under real-world conditions, where the complexities of hardware configurations and scalability are challenging to replicate in controlled environments.
Furthermore, the inherent variability of real-world data streams makes accurate simulation and benchmarking difficult.
Most studies to date are restricted to small-scale experiments, with limited exploration of how DSP frameworks perform on large-scale systems and High-Performance Computing (HPC) environments.
HPC systems utilize parallel processing and distributed computing on large-scale clusters to efficiently manage complex computational tasks, making them essential for real-time stream and batch processing of massive datasets across various domains.
It is therefore essential to evaluate DSP frameworks also on large-scale and HPC systems.
However, to the best of our knowledge, no suitable benchmarking tools currently exist for this purpose.
This highlights the need for developing more robust and scalable benchmarking suite to effectively bridge these gaps.

This work introduces \textit{SProBench}, a highly scalable and high-throughput stream processing benchmark specifically designed to seamlessly support large-scale HPC systems.
SProBench is open source\footnote{available from \url{https://github.com/apurvkulkarni7/SProBench}.} and provides native support for the Slurm batch management system.
SProBench is designed to generate high-velocity data streams, capable of producing millions of events per second.
It offers configurable parameters, including event size, data distribution, and frequency, enabling the simulation of realistic workloads to accurately assess the performance and scalability of any stream processing framework.

This work is organized as follows:
Section~\ref{sec:literature} reviews the current literature and explains how the proposed benchmark bridges the gaps.
Section~\ref{sec:methodology} provides a comprehensive explanation of the design of the benchmark, including its architecture, workloads, metrics.
Section~\ref{sec:experiment} demonstrates the functionality of the benchmark in a scalability experiments, and Section~\ref{sec:concludion} concludes this work providing future directions.

\section{Related Works}
\label{sec:literature}
Over the past decade, DSP benchmarking tools have undergone substantial advancements to address increasing data complexity and evolving computational requirements.
Most benchmarks focus on latency and throughput as the main metrics, often complementing them with another metric such as memory or CPU usage.
For effective benchmarking of DSP frameworks, the frequency, distribution and size of the generated workload should represent the real-world scenarios.

Earliest works like Linear Road Benchmark~\cite{arasuLinearRoadAStream2004} measured the performance of the streaming frameworks, such as AURORA and STREAM, with the focus on latency.
The data consisted of simulated traffic and sensor data for toll systems, with a workload consisting of aggregation and joining queries on a database.
%
%
Yahoo streaming benchmark (YSB)~\cite{chintapalliBenchmarkingStreamingComputation2016a} is another popular benchmark that focuses on metrics like throughput and latency to asses scalability of the streaming frameworks such as Apache Storm, Flink and Spark Streaming.
In YSB, the data consisted of synthetically generated advertisement campaign, on which, workloads such as mapping, filtering, transformation, joining and windowing were performed.
DSPBench~\cite{bordinDSPBenchSuiteBenchmark2020} which is a benchmark suite for distributed DSP systems, covers a wider range of application domains including finance, telecommunications, and sensor networks.
This benchmark employs metrics such as throughput, latency, memory, network, and CPU usage to measure the performance of Apache Storm and Spark Streaming frameworks.
Theodolite~\cite{henningTheodoliteScalabilityBenchmarking2021,henningBenchmarkingScalabilityStream2024} is a benchmarking method for evaluating the scalability of cloud-native applications, particularly stream processing within micro-services.
The benchmark consists of different processing pipelines providing the capability to test different operations of the framework and uses non-traditional approaches to measure the performance, namely using metrics such as required number of processing instances per workload.
Among enterprise benchmark suites, ESPBench~\cite{hesseESPBenchEnterpriseStream2021} is designed to evaluate DSPs using business and sensor data from manufacturing scenarios.
Its workload includes queries that cover core stream processing functionalities, and it provides a toolkit for data ingestion, result validation, and objective latency measurements.
The benchmark is used to compare systems like Apache Spark Streaming, Flink, and Hazelcast Jet, highlighting the importance of result validation.
OSPBench\cite{vandongenInfluencingFactorsScalability2021} benchmark evaluates the scalability of stream processing jobs in Flink, Kafka Streams, Spark Streaming, and Structured Streaming using national traffic data and two distinct pipelines, namely memory intensive and CPU intensive.
It analyses both horizontal and vertical scaling using metrics such as latency, throughput, scaling efficiency, CPU, garbage collection (GC), network IO, file-system and disk IO.
There are also more specific benchmark tools such as SPBench~\cite{garciaSPBenchFrameworkCreating2023}, which only supports C++ based frameworks like FastFlow.
To gauge the performance it relies on metrics like latency, throughput, CPU and memory usage, and consists of various workloads ranging from computer vision application like lane detection and person recognition to compression-decompression workloads.
While most benchmark suites support Linux based systems, to the best of our knowledge, none of them supports SLURM integration. 
Theodolite, built for cloud-native apps, may work with Kubernetes and integrate with SLURM via plugins or configurations.
However, this potential integration is not reported in the literature.
Table~\ref{tab:benchmark_comparison} shows a comparative overview of different DSP benchmark suites.

\begin{table}[h]
\centering
\small

    \begin{tabular}{|l|ccccc|ccccc|ccccccc|c|ccc|}
        \hline
        & \multicolumn{5}{c|}{Native Metrics} & \multicolumn{5}{c|}{Ext. Metrics} & \multicolumn{7}{c|}{DSP Framework} & & \multicolumn{3}{c|}{~~~~Support~~~~}
        \\ 
        \rotatebox[origin=l]{0}{Benchmarks} & 
        \rotatebox[origin=l]{90}{Latency} & 
        \rotatebox[origin=l]{90}{Throughput} & 
        \rotatebox[origin=l]{90}{CPU usage} & 
        \rotatebox[origin=l]{90}{Memory usage} &
        \rotatebox[origin=l]{90}{Garbage collection} &
        \rotatebox[origin=l]{90}{CPU mem. bandwidth} &
        \rotatebox[origin=l]{90}{Network usage} &
        \rotatebox[origin=l]{90}{filesystem read/write} &
        \rotatebox[origin=l]{90}{I/O} & 
        \rotatebox[origin=l]{90}{Energy} & 
        \rotatebox[origin=l]{90}{Aurora/Stream} & 
        \rotatebox[origin=l]{90}{Apache Storm} & 
        \rotatebox[origin=l]{90}{Apache Flink} & 
        \rotatebox[origin=l]{90}{Apache Spark} & 
        \rotatebox[origin=l]{90}{Apache Kafka} & 
        \rotatebox[origin=l]{90}{Hazelcast Jet} & 
        \rotatebox[origin=l]{90}{Fastflow/Windflow} & 
        \rotatebox[origin=l]{90}{Experiment Automation} &
        \rotatebox[origin=l]{90}{SLURM} & 
        \rotatebox[origin=l]{90}{Message Broker} & 
        \rotatebox[origin=l]{90}{Max Doc. Throughput}        
        \\ \hline
         Linear Road & $\bullet$  &           &            &           &           &           &           &           &           &           & $\bullet$ &           &           &           &           &           &           &           & &           & $0.1~M/s$ \\
         YSB         & $\bullet$  & $\bullet$ &            &           &           &           &           &           &           &           &           & $\bullet$ & $\bullet$ & $\bullet$ &           &           &           &           & &           & $0.2~M/s$ \\
         DSPBench    & $\bullet$  & $\bullet$ & $\bullet$  & $\bullet$ &           &           &           & $\bullet$ &           &           &           & $\bullet$ &           & $\bullet$ &           &           &           &           & & $\bullet$ & $0.8~M/s$ \\
         Theodolite  & $\bullet$  &           &            &           &           &           &           &           &           &           &           &           & $\bullet$ & $\bullet$ & $\bullet$ & $\bullet$ &           &           & & $\bullet$ & $1.0~M/s$   \\
         ESPBench    & $\bullet$  & $\bullet$ & $\bullet$  &           &           &           & $\bullet$ &           & $\bullet$ &           &           &           & $\bullet$ & $\bullet$ &           & $\bullet$ &           &           & & $\bullet$ & $0.1~M/s$    \\
         SPBench     & $\bullet$  & $\bullet$ & $\bullet$  & $\bullet$ &           &           &           &           &           &           &           &           &           &           &           &           & $\bullet$ &           & &           & $0.5~K/s$  \\
         OSPBench    & $\bullet$  & $\bullet$ & $\bullet$  & $\bullet$ & $\bullet$ & $\bullet$ &           &           &           &           &           &           & $\bullet$ & $\bullet$ & $\bullet$ &           &           &           & & $\bullet$ & $0.8~M/s$   \\
         \hline
         SProBench   & $\bullet$  & $\bullet$ & $\bullet$  & $\bullet$ & $\bullet$ & $\bullet$ & $\bullet$ & $\bullet$ & $\bullet$ & $\bullet$ &           & $\circ$ & $\bullet$ & $\bullet$ & $\bullet$ & $\circ$    & $\circ$  & $\bullet$ & $\bullet$ & $\bullet$ & \textbf{40~$M/s$} \\
         \hline
         \multicolumn{22}{l}{\hfill$\bullet$ - full support, $\circ$ - partial support}
\end{tabular}
    \caption{Comparison of data steam processing benchmarks}
    \label{tab:benchmark_comparison}
\end{table}%

Empirical results from running existing benchmark suites in our test environment revealed two key challenges: first, many benchmark suites demonstrate inefficient execution and cannot fully utilize available resources;
second, a considerable number of benchmarks are built on outdated frameworks, which have undergone significant updates and major version changes, making the original benchmarks less relevant and less effective for evaluating modern computing systems.
Additionally, we found that most benchmark suites struggle to scale beyond a certain throughput threshold.
While the scalability of benchmark suites is influenced by various parameters, our experiments observed a maximum throughput ranging from 1 to 4 million events, consistent with findings reported in the existing literature~\cite{henningBenchmarkingScalabilityStream2024,henningTheodoliteScalabilityBenchmarking2021}.
Although some benchmarks specified the size of each event~\cite{vandongenInfluencingFactorsScalability2021}, explicit evidence of throughput in terms of event size was absent in the majority of benchmarks.
The benchmarks reviewed in this work showed limited flexibility for testing individual components, such as the message broker and stream processor, and offered minimal support for custom message sizing.
Furthermore, they were not fully optimized to seamlessly support multiple experiments on HPC systems, particularly those utilizing SLURM.
These aspects introduced challenges in pinpointing bottlenecks within the pipeline and addressing inefficiencies during benchmark execution.
The SProBench benchmark suite introduced in this work addresses the previously mentioned limitations through several key improvements.
These include seamless integration with SLURM, customizable event sizing, automated support for running multiple concurrent experiments on SLURM, and enhanced flexibility for evaluating various components such as message queues and stream processing frameworks.
Additionally, SProBench's workload generator exhibits remarkable scalability, exceeding 20 million events per second and achieving a throughput of approximately 0.5 GB/second on a single node, demonstrating its ability to handle high-volume data generation effectively.

\section{SProBench}
\label{sec:methodology}
\begin{figure}[h]
    \centering
    \scalebox{0.9}{    \begin{tikzpicture}[
                inner sep=0mm, outer sep=0mm,
                component/.style= {rounded corners, align=center, draw},
                myarrow/.style= {-stealth,very thick},
            ]
        \newlength{\archImgHeight}\setlength{\archImgHeight}{45mm}

        \node(cli) [component,text width=0.1\textwidth, minimum width=0.1\textwidth, minimum height=\archImgHeight, align=center] {CLI Interface};
        \node(slurm) [component, right=5mm of cli.north east, anchor=north west, minimum width=0.45\textwidth, minimum height=0.15\archImgHeight] {SLURM Integration};
        \node(gen) [component, below=5mm of slurm, minimum width=0.45\textwidth, minimum height=0.15\archImgHeight] {Workload Generator};
        \node(brk) [component, below=5mm of gen, minimum width=0.45\textwidth, minimum height=0.15\archImgHeight] {Message Broker};
        \node(proc) [component, at=(cli.south-|brk), anchor=south, minimum width=0.45\textwidth, minimum height=0.2\archImgHeight] {Stream Processing Framework};
        \node(monitor) [component,text width=0.1\textwidth, minimum width=0.1\textwidth, minimum height=\archImgHeight, align=center,right=5mm of slurm.north east, anchor=north west] {Monit-oring};
        \node(storage) [component,minimum width=0.12\textwidth, minimum height=0.2\archImgHeight,right=5mm of monitor.north east, anchor=north west] {Storage};
        \node(postproc) [component,text width=0.12\textwidth,minimum width=0.12\textwidth, minimum height=0.2\archImgHeight,below=5mm of storage] {Offline Post Proc.};
        \node(extm) [component,text width=0.12\textwidth,minimum width=0.12\textwidth, minimum height=0.2\archImgHeight,at=(monitor.south-|storage), anchor=south] {External Metrics};

        \draw[myarrow] (cli.east|-slurm) -- (slurm);
        \draw[myarrow] (cli.east|-gen) -- (gen);
        \draw[myarrow] (cli.east|-brk) -- (brk);
        \draw[myarrow] (cli.east|-proc) -- (proc);
        \draw[myarrow] (gen) -- (brk);
        \node[at=(brk.south), xshift=-10mm] (brkcoordOne) {};
        \node[at=(brk.south), xshift=10mm] (brkcoordTwo) {};
        \draw[myarrow] (brkcoordOne|-brk.south) -- (brkcoordOne|-proc.north);
        \draw[myarrow] (brkcoordTwo|-proc.north) -- (brkcoordTwo|-brk.south);

        \draw[myarrow] (gen) -- (monitor.west|-gen);
        \draw[myarrow] (brk) -- (monitor.west|-brk);
        \draw[myarrow] (proc) -- (monitor.west|-proc);
        \draw[myarrow] (extm) -- (monitor.east|-extm);

        \draw[myarrow] (monitor.east|-storage) -- (storage);
        \draw[myarrow] (storage) -- (postproc);
    \end{tikzpicture}}
    \caption{Benchmark architecture}
    \label{fig:benchmark_architecture}
\end{figure}
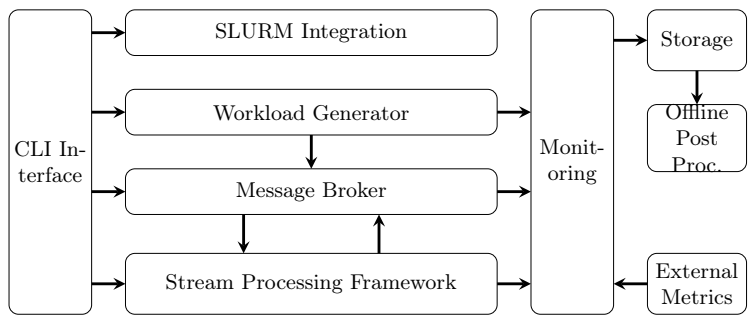

Stream processing benchmark suites often consist of 3 common components: \textit{workload generator} to simulate the data stream, \textit{monitoring unit} to collect relevant metrics, and \textit{post-processing unit} which aggregates and validates the monitoring data.
The flexibility and comprehensiveness of each of these components are crucial factors that significantly influence the overall performance, accuracy, and usability of each benchmark suite.
Additionally, the design of a practical benchmark suite demands several key factors, including user-friendly interface, centralized and flexible configurations, interoperable framework interface, high scalability, and automation.
The benchmark suite proposed in this work is designed based on a flexible modular architecture comprising independent components, which makes SProBench adaptable to virtually any cluster topology and DSP framework.
Figure~\ref{fig:benchmark_architecture} provides a schematic representation of SProBench's architecture, highlighting its key components.
The workload generator simulates various real-world workloads as well as custom workloads per user settings.
Message broker decouples the workload generator and stream processing layer.
The stream processing layer performs computation on the stream generated by workload generator, and consists of different stream processing application logic for different frameworks.
In the current implementation, Apache Flink, Apache Spark Streaming, and Apache Kafka Stream are fully integrated.
SProBench's well-defined interface and modular architecture enable seamless integration of any other DSP framework with minimal modifications.

All components are monitored using different metric collector tools such as throughput and latency collector, as well as java management extension (JMX) tool which collects metrics related to the java virtual machine.
Other system metrics like network and memory bandwidth are collected using external monitoring tools such as Pika~\cite{winklerAutomaticDetectionHPC2023} and MetricQ~\cite{ilscheMetricQScalableInfrastructure2019}.
The monitoring layer transmits all metrics to a central storage.
The stored metrics are then aggregated and validated by the post-processing unit and are utilized for further offline analysis.

In addition, SProBench provides a command-line interface (CLI) for the orchestration of all components, setting up frameworks, compiling the resources and performing the benchmarks.
The CLI enables out-of-the-box execution of benchmark suite on local machines as well as on SLURM-based clusters, supporting both interactive and batch executions.
CLI's internal workflow management utility provides fully automatic, reproducible, and scalable benchmarking experiments.
%
This interface also facilitates the allocation of resources in a SLURM-based environment.
By referencing the memory and CPU requirements specified in the configuration file, the interface automatically determines the appropriate SLURM job parameters.
Once the resources are allocated, the interface defines all the environment variables necessary for the benchmark processes.
%
%
A single configuration file serves as a master control point for setting up various options across all components in the benchmark.
This streamlined approach enables the execution of multiple experiments with ease, allowing for efficient testing of different scenarios.
For instance, by maintaining a consistent parallelism in the processing pipeline, it is possible to test multiple workloads without modifying or need of creating multiple configuration files.
This flexibility is achieved through a well-defined configuration file along with the powerful CLI interface, which can be tailored to accommodate diverse testing requirements and scenarios.
\begin{figure*}[h]
    \centering
    \begin{subfigure}[b]{0.33\textwidth}
        \centering
        \scalebox{0.45}{\begin{tikzpicture}[
        mainProcess/.style= {rounded corners, minimum width=30mm, minimum height=53mm, draw},
        subComponent/.style= {rounded corners, minimum width=28mm, minimum height=5mm, draw, fill=white},
        subsubComponent/.style= {rounded corners, minimum width=26mm, minimum height=5mm, draw,fill=white},
        container/.style = {draw, rectangle, dashed, inner sep=3mm},
        outer sep=0mm,
    ]
    \newlength{\vertSpacingSubCompOne}\setlength{\vertSpacingSubCompOne}{2mm}
    \newlength{\vertSpacingSubSubCompOne}\setlength{\vertSpacingSubSubCompOne}{1mm}
    
    \node(benchmarkDriver) [mainProcess, label={[yshift=-5mm]Benchmark Driver}, fill=blue!20] {};
    \node(dataGenerator) [subComponent, below=7mm of benchmarkDriver.north] {Data Generator};
    \node(kafkaBroker) [subComponent, below=\vertSpacingSubCompOne of dataGenerator] {Kafka Broker};
    \node(metricCollectors) [subComponent, below=\vertSpacingSubCompOne of kafkaBroker, minimum height=20mm, label={[yshift=-5mm]Metric Collectors}] {};
    \node(metricCollecDataGen) [subsubComponent,below=5mm of metricCollectors.north] {Data Generator};
    \node(metricCollecKafkaBroker) [subsubComponent,below=\vertSpacingSubCompOne of metricCollecDataGen] {Kafka broker};
    \node(masterProcess) [subComponent, below=\vertSpacingSubCompOne of metricCollectors] {Master Process};
    
    \node(worker) [mainProcess, right=5mm of benchmarkDriver, label={[yshift=-5mm]Worker}, fill=green!20] {};
    \node(workerProcesses) [subComponent, below=7mm of worker.north] {Worker processes};
    \node(tasks) [subComponent, below=\vertSpacingSubCompOne of workerProcesses, minimum height=30mm, label={[yshift=-5mm]Tasks}] {};
    \node(taskOne) [subsubComponent, below=5mm of tasks.north] {Parallel Task 1};
    \node(taskTwo) [subsubComponent, below=\vertSpacingSubSubCompOne of taskOne] {Parallel Task 2};
    \node(taskX) [subsubComponent, below=\vertSpacingSubSubCompOne of taskTwo] {...};
    \node(taskN) [subsubComponent, below=\vertSpacingSubSubCompOne of taskX] {Parallel Task N};
    \node(metricCollectorWorker) [subComponent, below=\vertSpacingSubCompOne of tasks] {Metric Collector};        

    \node[container, fit=(benchmarkDriver) (worker), label={[yshift=1mm]Node$_1$}] (node1) {};
    
    \draw[stealth-stealth, very thick] (benchmarkDriver) -- (worker);
    
    \node(scaleupArrowTop) [right=2mm of taskOne.north east] {};
    \node(scaleupArrowDown) [right=2mm of taskN.south east] {};
    \draw[stealth-stealth, very thick] (scaleupArrowTop) -- node [anchor=north,midway,right=1.5mm, rotate=90, yshift=-2.5mm,xshift=-15mm]{Scaling across CPUs} (scaleupArrowDown);
    
\end{tikzpicture}}
        \caption{Scale-Up single-node}
        \label{fig:benchmark_setup_scaleup_a}
    \end{subfigure}%
    ~~%
    \begin{subfigure}[b]{0.33\textwidth}
        \centering
        \scalebox{0.45}{\begin{tikzpicture}[
        mainProcess/.style= {rounded corners, minimum width=30mm, minimum height=53mm, draw},
        subComponent/.style= {rounded corners, minimum width=28mm, minimum height=5mm, draw, fill=white},
        subsubComponent/.style= {rounded corners, minimum width=26mm, minimum height=5mm, draw,fill=white},
        container/.style = {draw, rectangle, dashed, inner sep=3mm},
    ]
    \newlength{\vertSpacingSubComp}\setlength{\vertSpacingSubComp}{2mm}
    \newlength{\vertSpacingSubSubComp}\setlength{\vertSpacingSubSubComp}{1mm}
    
    \node(benchmarkDriver) [mainProcess, label={[yshift=-5mm]Benchmark Driver}, fill=blue!20] {};
    \node(dataGenerator) [subComponent, below=7mm of benchmarkDriver.north] {Data Generator};
    \node(kafkaBroker) [subComponent, below=\vertSpacingSubComp of dataGenerator] {Kafka Broker};
    \node(metricCollectors) [subComponent, below=\vertSpacingSubComp of kafkaBroker, minimum height=20mm, label={[yshift=-5mm]Metric Collectors}] {};
    \node(metricCollecDataGen) [subsubComponent,below=5mm of metricCollectors.north] {Data Generator};
    \node(metricCollecKafkaBroker) [subsubComponent,below=\vertSpacingSubComp of metricCollecDataGen] {Kafka broker};
    \node(masterProcess) [subComponent, below=\vertSpacingSubComp of metricCollectors] {Master Process};
    \node[container, fit=(benchmarkDriver) (benchmarkDriver), label={[yshift=1mm]Node$_1$}] (node1) {};

    \node(worker) [mainProcess, right=10mm of benchmarkDriver, label={[yshift=-5mm]Worker}, fill=green!20] {};
    \node(workerProcesses) [subComponent, below=7mm of worker.north] {Worker processes};
    \node(tasks) [subComponent, below=\vertSpacingSubComp of workerProcesses, minimum height=30mm, label={[yshift=-5mm]Tasks}] {};
    \node(taskOne) [subsubComponent, below=5mm of tasks.north] {Parallel Task 1};
    \node(taskTwo) [subsubComponent, below=\vertSpacingSubSubComp of taskOne] {Parallel Task 2};
    \node(taskX) [subsubComponent, below=\vertSpacingSubSubComp of taskTwo] {...};
    \node(taskN) [subsubComponent, below=\vertSpacingSubSubComp of taskX] {Parallel Task N};
    \node(metricCollectorWorker) [subComponent, below=\vertSpacingSubComp of tasks] {Metric Collector};        
    \node[container, fit=(worker) (worker), label={[yshift=1mm]Node$_2$}] (node2) {};

    \draw[stealth-stealth, very thick] (node1) -- (node2);
    
    \node(scaleupArrowTop) [right=2mm of taskOne.north east] {};
    \node(scaleupArrowDown) [right=2mm of taskN.south east] {};
    \draw[stealth-stealth, very thick] (scaleupArrowTop) -- node [anchor=north,midway,right=1.5mm, rotate=90, yshift=-2.5mm,xshift=-15mm]{Scaling across CPUs} (scaleupArrowDown);

    
\end{tikzpicture}}\\
        \caption{Scale-Up multi-node}
        \label{fig:benchmark_setup_scaleup_b}
    \end{subfigure}
    \begin{subfigure}[b]{0.3\textwidth}
        \centering
        \scalebox{0.45}{\begin{tikzpicture}[
    mainProcess/.style= {rounded corners, minimum width=30mm, minimum height=50mm, draw},
    subComponent/.style= {rounded corners, minimum width=28mm, minimum height=5mm, draw, fill=white},
    subsubComponent/.style= {rounded corners, minimum width=26mm, minimum height=5mm, draw,fill=white},
    container/.style = {draw, rectangle, dashed, inner sep=3mm},
]
\setlength{\vertSpacingSubComp}{1mm}
\setlength{\vertSpacingSubSubComp}{1mm}

\node(benchmarkDriver) [mainProcess, label={[yshift=-5mm]Benchmark Driver}, fill=blue!20] {};
\node(dataGenerator) [subComponent, below=5mm of benchmarkDriver.north] {Data Generator};
\node(kafkaBroker) [subComponent, below=\vertSpacingSubComp of dataGenerator] {Kafka Broker};
\node(metricCollectors) [subComponent, below=\vertSpacingSubComp of kafkaBroker, minimum height=20mm, label={[yshift=-5mm]Metric Collectors}] {};
\node(metricCollecDataGen) [subsubComponent,below=5mm of metricCollectors.north] {Data Generator};
\node(metricCollecKafkaBroker) [subsubComponent,below=\vertSpacingSubComp of metricCollecDataGen] {Kafka broker};
\node(masterProcess) [subComponent, below=\vertSpacingSubComp of metricCollectors] {Master Process};
\node[container, fit=(benchmarkDriver) (benchmarkDriver), label={[yshift=1mm]Node$_1$}] (node1) {};

\node(worker) [mainProcess, right=15mm of benchmarkDriver, fill=white, draw=white] {};       
\node[container, fit=(worker) (worker), label={[yshift=0mm]Node$_N$}] (nodeN) {};
\node(worker) [mainProcess, right=18mm of benchmarkDriver, yshift=-5mm, fill=white, draw=white] {};       
\node[container, fit=(worker) (worker), label={[yshift=0mm]Node$_{N-1}$}] (nodeN1) {};
\node(worker) [mainProcess, right=21mm of benchmarkDriver, yshift=-10mm, fill=white, draw=white] {};       
\node[container, fit=(worker) (worker), label={[yshift=0mm]Node$...$}] (nodeN2) {};
\node(worker) [mainProcess, right=24mm of benchmarkDriver, yshift=-15mm, label={[yshift=-5mm]Worker}, fill=green!20] {};
\node(workerProcesses) [subComponent, below=5mm of worker.north] {Worker processes};
\node(tasks) [subComponent, below=\vertSpacingSubComp of workerProcesses, minimum height=30mm, label={[yshift=-5mm]Tasks}] {};
\node(taskOne) [subsubComponent, below=5mm of tasks.north] {Parallel Task 1};
\node(taskTwo) [subsubComponent, below=\vertSpacingSubSubComp of taskOne] {Parallel Task 2};
\node(taskX) [subsubComponent, below=\vertSpacingSubSubComp of taskTwo] {...};
\node(taskN) [subsubComponent, below=\vertSpacingSubSubComp of taskX] {Parallel Task N};
\node(metricCollectorWorker) [subComponent, below=\vertSpacingSubComp of tasks] {Metric Collector};        
\node[container, fit=(worker) (worker), label={[yshift=0mm]Node$_2$}] (node2) {};

\draw[stealth-stealth, very thick] (node1.east) -- (node2.west);
\draw[stealth-stealth, very thick] (node1.east) -- (nodeN2.west);
\draw[stealth-stealth, very thick] (node1.east) -- (nodeN1.west);
\draw[stealth-stealth, very thick] (node1.east) -- (nodeN.west);

\node(scaleupArrowTop) [right=1mm of nodeN.north east] {};
\node(scaleupArrowDown) [right=2mm of node2.north east] {};
\draw[stealth-stealth, very thick] (scaleupArrowTop) -- node [text width=20mm,midway,shift={(14mm,2mm)}]{Scaling across Nodes} (scaleupArrowDown);

\end{tikzpicture}}
        \caption{Scale-Out}
        \label{fig:benchmark_setup_scaleout}
    \end{subfigure}
    \caption{Benchmark process setup for scale-up and scale-out experimentation}
    \label{fig:benchmark_setup_scaleup}
\end{figure*}
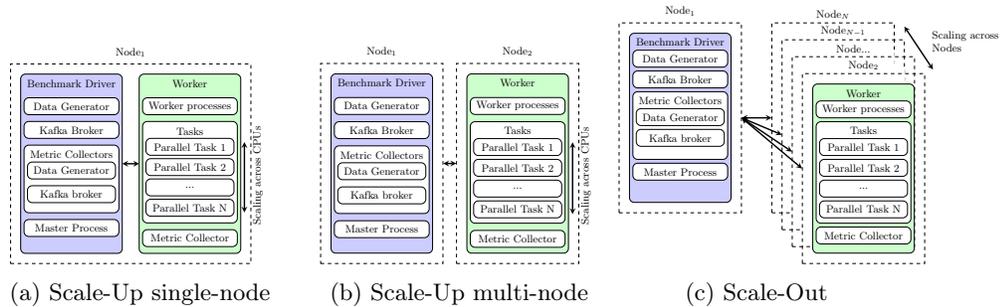


SProBench provides flexibility in grouping and organizing its components across various setups.
These setups enable different levels of isolation between the benchmark driver and worker components.
Figure~\ref{fig:benchmark_setup_scaleup} illustrate several example scenarios of scale-up and scale-out experiments, respectively.

\subsection{Benchmarking Workflow}\label{subsec:workflow}

Workflow management in large-scale benchmarking is challenging due to the complexity of coordinating processes, datasets, and resources.
Scaling workflows becomes harder as benchmarks grow, requiring efficient task management and resource utilization.
Inconsistent workflows and poor documentation of setups, such as missing details on software versions or configurations, hinder reproducibility.
To address these issues, the SProBench workflow management system logs every step of an experiment for traceability.
It automates most benchmarking tasks, reduces human error, and ensures consistency across experiments.

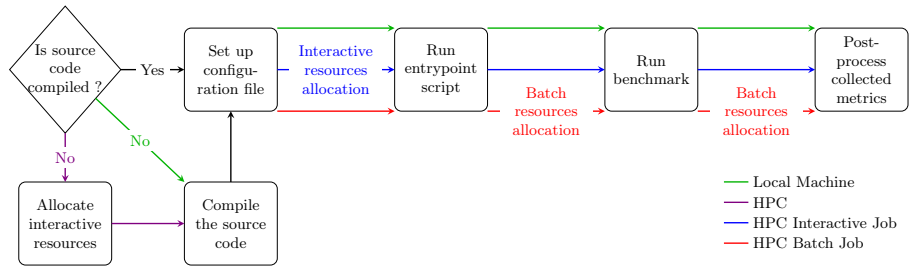
\begin{figure}[h]
    \centering
    \scalebox{0.65}{\begin{tikzpicture}[
    node distance=20mm,
    boxStyle/.style= {rounded corners,minimum height=17mm, draw, text width=17mm, align=center},
    arrowStyle/.style= {-stealth, thick},
    arrowStyleLocalmachine/.style= {arrowStyle, color=green!70!black},
    arrowStyleHPC/.style= {arrowStyle, color=blue!50!red},
    arrowStyleInteractive/.style= {arrowStyle, color=blue},
    arrowStyleBatch/.style= {arrowStyle, color=red},
    decision/.style= {diamond,minimum width=7em,minimum height=8em, text centered, draw},
    inner sep=1mm, outer sep=0mm
    ]
    \node (runType) [decision] {};
    \node (runTypeText) [at=(runType), text width=5em, align=center] {Is source code\\compiled ?};
    \node (interRes) [boxStyle, below= of runType, yshift=10mm] {Allocate interactive resources};
    \node (config) [boxStyle, right= of runType, xshift=-7mm] {Set up configuration file};
    \node (compile) [boxStyle, at=(interRes-|config)] {Compile the source code};
    \node (run) [boxStyle, right= of config, xshift=4mm] {Run entrypoint script};
    \node (run-benchmark) [boxStyle, right= of run, xshift=4mm] {Run benchmark};
    \node (post-process) [boxStyle, right= of run-benchmark, xshift=4mm] {Post-process collected metrics};

    \draw[arrowStyle] (runType) -- (config) node[midway, fill=white, align=center]{Yes};
    \draw[arrowStyleHPC] (runType) -- (interRes) node[midway, align=center, fill=white]{No};
    \draw[arrowStyleLocalmachine] (runType) -- (compile.north west) node[midway, align=center, fill=white]{No};
    \draw[arrowStyleHPC] (interRes) -- (compile);
    \draw[arrowStyle] (compile) -- (config);
    \draw[arrowStyleLocalmachine] (config.north east) -- (run.north west);
    \draw[arrowStyleLocalmachine] (run.north east) -- (run-benchmark.north west);
    \draw[arrowStyleLocalmachine] (run-benchmark.north east) -- (post-process.north west);
    \draw[arrowStyleInteractive] (config) -- node[midway, text width=17mm, fill=white, align=center]{Interactive resources allocation} (run);
    \draw[arrowStyleInteractive] (run) -- (run-benchmark);
    \draw[arrowStyleInteractive] (run-benchmark) -- (post-process);
    \draw[arrowStyleBatch] (config.south east) -- (run.south west);
    \draw[arrowStyleBatch] (run.south east) -- node[midway, text width=17mm, fill=white, align=center]{Batch resources \\ allocation} (run-benchmark.south west);
    \draw[arrowStyleBatch] (run-benchmark.south east) --node[midway, text width=17mm, fill=white, align=center]{Batch resources \\ allocation} (post-process.south west);
    
    \node(legendLocal) [at=(current bounding box.south east),anchor=south east, yshift=15mm, xshift=-10mm]  {Local Machine};
    \draw[arrowStyleLocalmachine,-] (legendLocal.west) -- +(-5mm,0mm);
    \node(legendHPC) [below=0mm of legendLocal.south west, anchor=north west]  {HPC};
    \draw[arrowStyleHPC,-] (legendHPC.west) -- +(-5mm,0mm);
    \node(legendInter) [below=0mm of legendHPC.south west, anchor=north west]  {HPC Interactive Job};
    \draw[arrowStyleInteractive,-] (legendInter.west) -- +(-5mm,0mm);
    \node(legendBatch) [below=0mm of legendInter.south west, anchor=north west]  {HPC Batch Job};
    \draw[arrowStyleBatch,-] (legendBatch.west) -- +(-5mm,0mm);
\end{tikzpicture}}
    \caption{Benchmark Workflow}
    \label{fig:benchmark_workflow}
\end{figure}

The benchmarking workflow via SProBench begins with obtaining the SProBench code repository and compiling the benchmark suite for the target environment.
Afterwards, the benchmark parameters need to be adjusted in the central configuration file; which is the only manual step of the benchmarking workflow.
In this configuration file, parameters such as workload, number of nodes and CPUs, degree of parallelism, memory, and so forth are defined.
Upon setting the configuration, the benchmark can be executed using the provided entrypoint script and relevant flags.
The script then identifies the target environment and whether it operates in interactive or batch mode. 
In case of an interactive job, the script verifies that sufficient resources are allocated according to the configuration file and then initiates the benchmark.
Conversely, for batch jobs the script calculates the required resources and submits a batch job request, which runs in the background and executes the benchmark within it.
Once the benchmarking process commences, the required directory structure is created, followed by the initiation of necessary processes and the processing pipeline.
Figure~\ref{fig:benchmark_workflow} illustrates an abstract overview of the benchmarking workflow.

Once the benchmarking process is complete, the post-processing scripts are executed to process the collected metrics.
The benchmark suite allows multiple experiments to be run from a single configuration file. , either with different configurations or the same configuration.
This enables simultaneous benchmarking e.g., with various workloads of 5M and 10M events, or multiple runs by the same workload. 
In addition, the transparent handling of parallel batch job execution and job dependencies is ensured.

\subsection{Workload Generation}\label{subsec:workload}

SProBench's workload generator produces synthetic data streams similar to real-world scenarios.
This multi-threaded Java application can produce up to 500,000 events per second per instance.
For increased throughput, multiple workload generators can operate in parallel.
The workload generator automatically adjusts the number of generators based on the requested total load specified in the configuration, thereby eliminating the need for users to manually manage this aspect.

The workload generator offers extensive customization options, allowing users to tailor characteristics such as frequency, throughput, and record size to create any desired workload.
By default, SProBench's workload generator produces synthetic streams of sensor data.
Each generated event follows a JSON format, containing a timestamp, sensor ID, and temperature value. The workload generator supports \textit{constant}, \textit{random}, and \textit{burst} generation patterns.
In constant mode, it produces events at a fixed frequency, whereas in random mode it generates events at a variable rate.
The random workload generation rate is subject to constraints such as minimum and maximum pauses between data generation and minimum and maximum frequencies of data generation.
In contrast, the burst mode produces data in bursts at a specified interval, with a desired workload frequency.
Notably, the burst mode can be considered a special case of the random interval generation, where the minimum and maximum pauses between data generation are the same, and the data generation frequency is constant.
The configuration file allows users to further fine-tune the workload generation by specifying parameters such as memory and CPU usage.
Additionally, the workload generator has the capability to set the size of each generated event, with the minimum event size being 27 bytes.

\subsection{Processing Pipeline}\label{subsec:processing_pipeline}

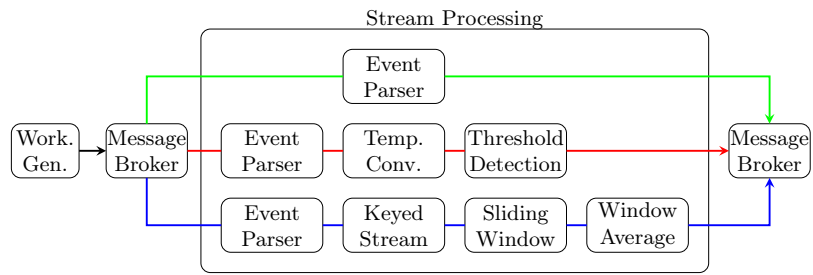
\begin{figure}[h]
    \centering
    \scalebox{0.9}{\begin{tikzpicture}[
        inner sep=0mm, outer sep=0mm,
        process/.style= {rounded corners, minimum width=18mm, minimum height=8mm, draw, text width=15mm, align=center},
        subProcess/.style= {rounded corners, minimum width=10mm, minimum height=8mm, draw, fill=white, text width=15mm, align=center},
    ]
    \node(gen) [process, text width=10mm,minimum width=10mm] { Work. Gen.};
    \node(msgbrk) [process,right=4mm of gen,text width=12mm,minimum width=10mm] {Message Broker};

    \node(eventParserP1)[subProcess, right=5mm of msgbrk.east]{Event Parser};
    \node(temperatureConvert)[subProcess, right=3mm of eventParserP1]{Temp. Conv.};
    \node(threasholdDetection)[subProcess, right=3mm of temperatureConvert]{Threshold Detection};

    \node(passthroughP0)[subProcess, above=3mm of temperatureConvert]{Event Parser};

    \node(P2eventParser)[subProcess, below=3mm of eventParserP1.south]{Event Parser};
    \node(P2keyed)[subProcess, right=3mm of P2eventParser]{Keyed Stream};
    \node(P2window)[subProcess, right=3mm of P2keyed]{Sliding Window};
    \node(P2avg)[subProcess, right=3mm of P2window]{Window Average};
    
    \node(streamproc) [draw, rounded corners, fit=(eventParserP1.west|-passthroughP0.north) (P2avg), inner sep=3mm, label={Stream Processing}] {};
    \node(msgbrkTwo) [process,right=3mm of streamproc,text width=12mm,minimum width=10mm] {Message Broker};

    \draw[-stealth,draw,thick] (gen) -- (msgbrk);
    \draw[-stealth,draw=green,thick] (msgbrk) -- (msgbrk|-passthroughP0) -- (passthroughP0) -- (passthroughP0-|msgbrkTwo) -- (msgbrkTwo);
    \draw[-stealth,draw=red,thick] (msgbrk) -- (eventParserP1) -- (temperatureConvert) -- (threasholdDetection) -- (msgbrkTwo);
    \draw[-stealth,draw=blue,thick] (msgbrk) -- (msgbrk |-P2eventParser)-- (P2eventParser) -- (P2keyed) -- (P2window) -- (P2avg) -- (P2avg-|msgbrkTwo) -- (msgbrkTwo);

\end{tikzpicture}}
    \caption{Processing Pipline}
    \label{fig:proccessing_pipeline}
\end{figure}

Based on insights from the literature and empirical analysis, this work covers three distinct classes of processing pipelines namely: \textit{pass-through}, \textit{CPU-intensive}, and \textit{memory-intensive}~\cite{henningTheodoliteScalabilityBenchmarking2021,vandongenInfluencingFactorsScalability2021}.
Figure~\ref{fig:proccessing_pipeline} depicts an overview of different processing pipelines.
%
Message brokers are positioned at both ends of the processing pipelines, serving as data queues for the incoming and outgoing data streams.
As shown in Figure~\ref{fig:proccessing_pipeline}, before pushing the data into any of the processing pipelines, the workload generator sends the data to the message broker; in this example Apache Kafka.
The left-hand message broker serves as the ingestion source, where raw data streams originate.
Message broker on the right-hand serves as the egestion target, where processed data streams are received.
The central section of the pipeline consists of stream processing engines to process the incoming data streams using frameworks like Apache Flink, Apache Spark Streaming or Kafka Stream.
SProBench defines three processing pipelines for each framework: pass-through, CPU-intensive, and memory-intensive.
These pipelines allow users to thoroughly compare how the same processing logic is executed across different SDP frameworks.
It is worth noting that these predefined pipelines are designed based on extensive experimentation, but users can also define custom processing logic tailored to their specific benchmarking objectives with minimal modifications.

The pass-through pipeline, depicted by the green line in Figure~\ref{fig:proccessing_pipeline}, is defined as baseline for evaluating the performance of the benchmark suite and the target system.
In this configuration, the generated data is transmitted through the message broker, ingested by the streaming engines, and then forwarded to the message broker without undergoing any processing.
CPU-Intensive pipeline, shown by the red line in the Figure~\ref{fig:proccessing_pipeline}, is designed to perform computationally intensive tasks, resulting in elevated CPU utilization.
As explained in previous section, the default workload in SProBench consists of synthesized sensor data.
In agreement with the default workload, the CPU-Intensive pipeline consists of various transformation operations.
These operations are utilized to perform a range of tasks, including parsing incoming sensor data into a tuple and converting the incoming sensor temperature data to degrees Fahrenheit, which then is checked against a certain threshold.
Memory-intensive pipeline is designed to assess the capability of stream processing frameworks in handling stateful operations.
This pipeline includes transformation processes that parse incoming sensor data.
The data stream generated by default workload generator in this work is keyed by the sensor ID.
As illustrated by the blue line in Figure~\ref{fig:proccessing_pipeline}, a sliding window is applied to calculate the average temperature for each sensor ID over the runtime period.
The calculated mean temperature for each sensor ID is then maintained as part of the operation's state.


\subsection{Metric Collection}\label{subsec:metrics}

There are numerous metrics that can be considered to evaluate DSP framework's performance in conjunction with system performance.
As discussed in Section~\ref{sec:literature}, many benchmark suites consider throughput and latency for their evaluations, along with CPU and memory usage.
In addition, metrics and properties such as resource efficiency, scalability, fault tolerance and time (runtime/startup) have been utilized in stream processing benchmarks~\cite{vogelSystematicMappingPerformance2023}.
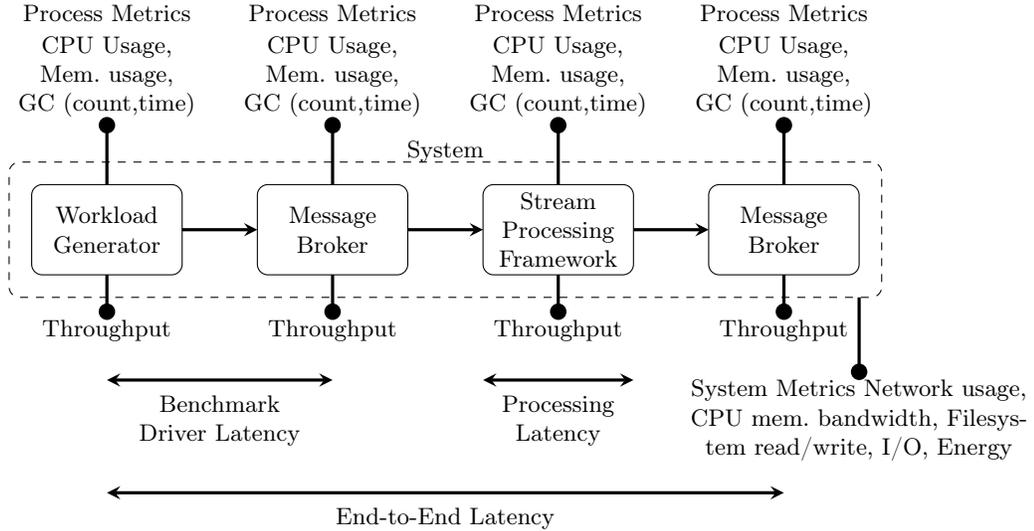
\begin{figure}[h]
    \centering
       \begin{tikzpicture}[
                inner sep=0mm, outer sep=0mm,
                component/.style= {rounded corners, align=center, draw, text width=20mm, minimum width=20mm, minimum height=12mm},
                myarrow/.style= {-stealth,very thick},
                metricarrow/.style= {-Circle[],very thick},
                processMetricsText/.style={text width=25mm,label={[yshift=2mm]Process Metrics},font=\footnotesize,align=center}
            ]
        \node(gen)[component]{Workload Generator};
        \node(brk)[component, right=10mm of gen]{Message Broker};
        \node(proc)[component, right=10mm of brk]{Stream Processing Framework};
        \node(brkTwo)[component, right=10mm of proc]{Message Broker};
        \node(sys)[component, fit=(gen) (brkTwo),dashed, inner sep=3mm, label={System}]{};

        \draw[myarrow] (gen) -- (brk);
        \draw[myarrow] (brk) -- (proc);
        \draw[myarrow] (proc) -- (brkTwo);

        \node(upMetrics) [at=(gen), yshift=15mm] {};
        \node(downMetrics) [at=(gen), yshift=-20mm] {};
        \node(downTgpMetrics) [at=(gen), yshift=-12mm] {};
        
        \draw[metricarrow] (gen) -- (upMetrics-|gen)node[processMetricsText,above] {CPU Usage, Mem. usage,\\GC (count,time)};
        \draw[metricarrow] (brk) -- (upMetrics-|brk)node[processMetricsText, above] {CPU Usage, Mem. usage,\\GC (count,time)};
        \draw[metricarrow] (proc) -- (upMetrics-|proc)node[processMetricsText, above] {CPU Usage, Mem. usage,\\GC (count,time)};
        \draw[metricarrow] (brkTwo) -- (upMetrics-|brkTwo)node[processMetricsText, above] {CPU Usage, Mem. usage,\\GC (count,time)};
        \draw[metricarrow] (sys.south-|brkTwo.east) -- (downMetrics-|brkTwo.east)node[below,text width=45mm, align=center] {System Metrics \footnotesize Network usage, CPU mem. bandwidth, Filesystem read/write, I/O, Energy};

        \draw[metricarrow] (gen) -- (downTgpMetrics-|gen)node[below] {Throughput};
        \draw[metricarrow] (brk) -- (downTgpMetrics-|brk)node[below] {Throughput};
        \draw[metricarrow] (proc) -- (downTgpMetrics-|proc)node[below] {Throughput};
        \draw[metricarrow] (brkTwo) -- (downTgpMetrics-|brkTwo)node[below] {Throughput};

        \node(driverLtyLeft) [at=(gen),yshift=-20mm]{};
        \node(endLtyLeft) [at=(gen),yshift=-35mm]{};
        \draw[stealth-stealth, very thick] (driverLtyLeft-|gen) -- (driverLtyLeft-|brk)node[midway,below, text width=25mm, yshift=-2mm, align=center]{Benchmark Driver Latency};
        \draw[stealth-stealth, very thick] (driverLtyLeft-|proc.west) -- (driverLtyLeft-|proc.east)node[midway,below, text width=25mm, yshift=-2mm, align=center]{Processing Latency};
        \draw[stealth-stealth, very thick] (endLtyLeft-|gen) -- (endLtyLeft-|brkTwo)node[midway,below, yshift=-2mm]{End-to-End Latency};
    \end{tikzpicture}
    \caption{Metrics Monitoring and Collection}
    \label{fig:metric_collection}
\end{figure}
Inspired by previous works, SProBench utilizes throughput, latency, CPU usage, memory usage, garbage collection (count and time), network usage, I/O, and energy consumption to evaluate the performance of stream processing frameworks and the underlying systems.
Metrics such as throughput are collected in terms of processed events per second, as well as in terms of size, i.e., processed size in Megabytes per second.
Throughput and latency are measured at several locations, as shown in Figure~\ref{fig:metric_collection}.
Latencies measured at different locations help in gathering information on different scales such as message benchmark driver latency, processing latency, and end-to-end latency, which in turn facilitates the identification of bottlenecks in each pipeline.

In addition to metrics related to data processing, SProBench also monitors and collects process metrics including memory usage and garbage collection (time and count).
As described earlier, to collect metrics from JVM-based processes, a Java based application is designed that relies on the JMX API~\cite{oracleJavaManagement} to gather all process metrics.
Furthermore, SProBench also monitors the target system's performance using external monitoring facilities.
Metrics collected in this step are highly dependent on the target system and its available monitoring mechanisms.
In the experiment described in Section~\ref{sec:experiment}, MetricQ~\cite{ilscheMetricQScalableInfrastructure2019} was used to collect energy consumption data of the underlying system, and other system metrics including CPU usage, system usage (memory bandwidth, FLOP, instructions per cycle, filesystem read/write), and network usage were collected using Pika~\cite{winklerAutomaticDetectionHPC2023}.





\section{Experiment}
\label{sec:experiment}

This section presents the performance of SProBench via experimental results.
Initially, experiments focus on scaling the benchmark with the Workload generator and Message Broker (Apache Kafka), demonstrating extreme scaling as presented in Table~\ref{tab:benchmark_comparison}.
Subsequently, scale-up experiment is conducted using the full processing pipeline, which involves the Workload generator, Message Broker (Apache Kafka), and Stream processing framework (Apache Flink).
%
%
For the proposed experiments Barnard HPC cluster powered by Red Hat Enterprise Linux is used. 
%
Barnard features 630 nodes, each equipped with dual Intel Xeon Platinum 8470 CPUs, delivering 104 cores per node.
Additionally, each node has 512 GB of RAM, implemented as 16 DDR5 memory modules operating at 4800 MT/s.
This configuration yields a total of 65,520 computational cores, making it suitable for large-scale parallel computing tasks and data-intensive research applications.
For the following experiments, a maximum of 200 GB of main memory is allocated for workers, with approximately 2 GB of heap memory allocated for each workload generator and 5 GB for Kafka, with 20 threads for I/O and 10 threads for network operations.
Each event has a size of 27 bytes.

First experiment investigates the workload scaling behavior in a simplified setup consisting solely of a workload generator and a Kafka message broker.
The configuration follows the setup described earlier, with the addition of 4 Kafka topic partitions and input workloads generated at rates of up to 0.5 million events per second.
The goal is to evaluate the scalability of both the workload generator and the Kafka broker, focusing on determining the maximum achievable throughput using multiple parallel generators in this streamlined configuration.

\begin{figure}
    \centering
    \begin{subfigure}[t]{0.33\textwidth}
        \centering
        \includegraphics[width=\textwidth]{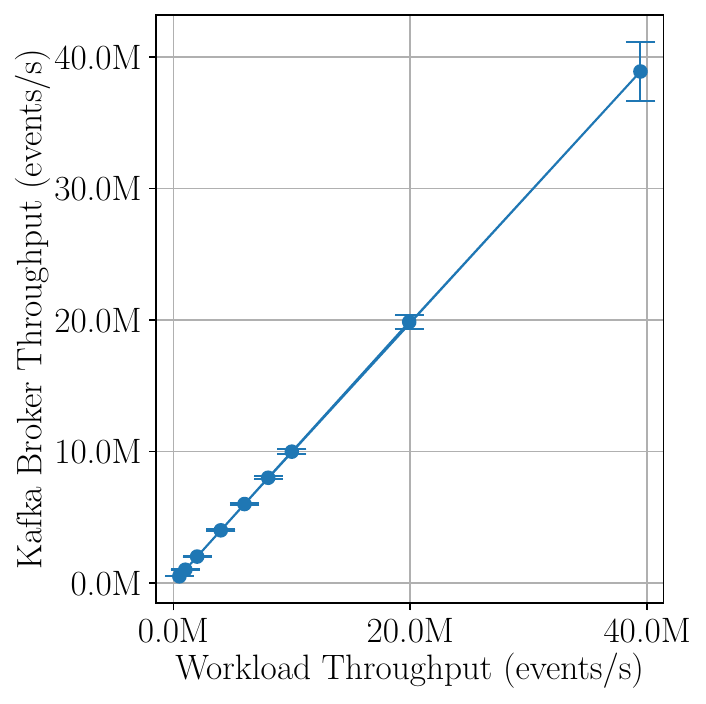}
        \label{fig:scaleup_gen_kafka_a}
    \end{subfigure}~
    \begin{subfigure}[t]{0.33\textwidth}
        \centering
        \includegraphics[width=\textwidth]{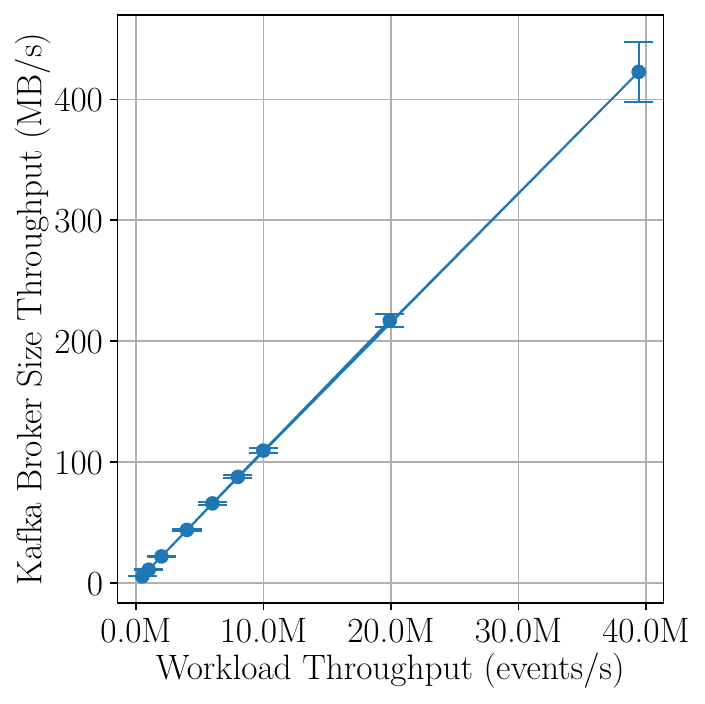}
        \label{fig:scaleup_gen_kafka_b}
    \end{subfigure}~
    \begin{subfigure}[t]{0.33\textwidth}
        \centering
        \includegraphics[width=\textwidth]{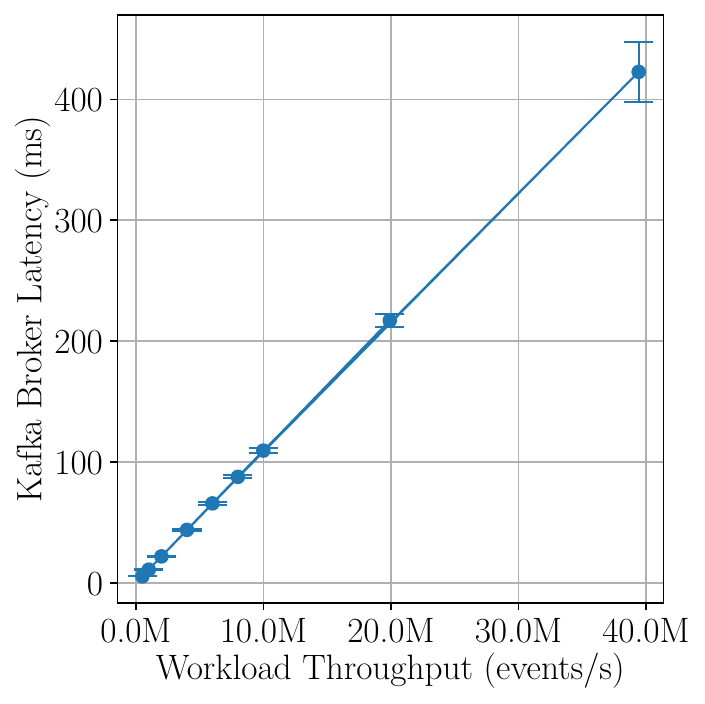}
        \label{fig:scaleup_gen_kafka_c}
    \end{subfigure}
    \caption{Scaling performance of Workload generator - Message Broker setup}
    \label{fig:gen_kafka_scaleup}
\end{figure}

Figure~\ref{fig:gen_kafka_scaleup} illustrates the linear scaling behavior of a Kafka broker system as workload throughput increases.
The results show a consistent 1:1 relationship between the broker system's throughput and the workload generator output.
The broker latency exhibits a similar linear scaling pattern as the workload intensifies.

\begin{figure}
    \centering
    \begin{subfigure}[t]{0.33\textwidth}
        \centering
        \includegraphics[width=\textwidth]{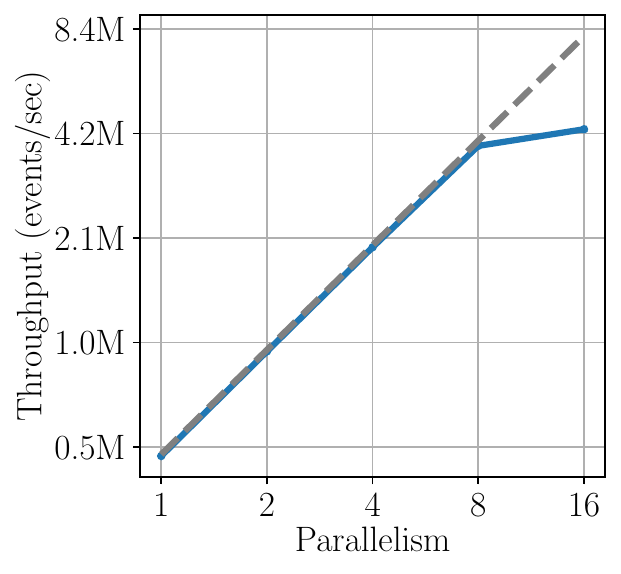}
        \caption{Parallel. Vs Throughput}
        \label{fig:scaleup_res_a}
    \end{subfigure}~
    \begin{subfigure}[t]{0.33\textwidth}
        \centering
        \includegraphics[width=\textwidth]{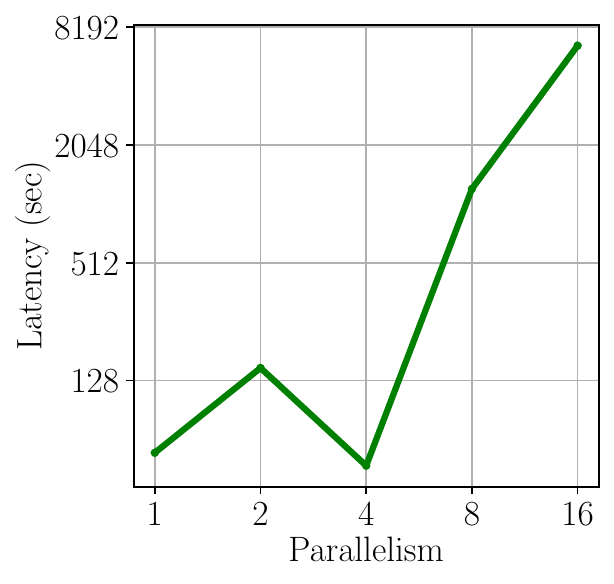}
        \caption{Parallelism Vs Latency}
        \label{fig:scaleup_res_b}
    \end{subfigure}~
    \begin{subfigure}[t]{0.33\textwidth}
        \centering
        \includegraphics[width=\textwidth]{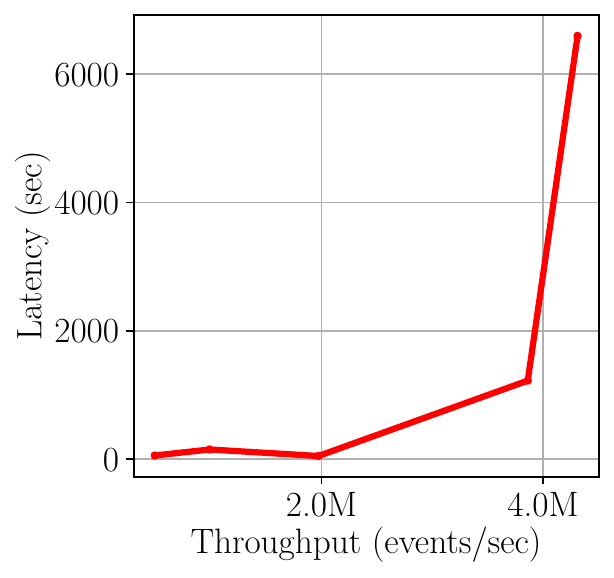}
        \caption{Parallelism Vs Latency}
        \label{fig:scaleup_res_b}
    \end{subfigure}
    \caption{Parallelism Vs Throughput and Latency}
    \label{fig:scaleup_res}
\end{figure}

Second experiment showcases the benchmark's ability to manage workloads with full process utilization, employing resources at varying levels of parallelism (1, 2, 4, 8, and 16 cores).
A CPU-intensive pipeline is used to showcase the benchmark's workings, with a constant workload frequency ranging from 0.5 million to 8 million events per second.
Figure~\ref{fig:scaleup_res} illustrates the benchmark's performance with varying counts of parallelism.
The framework demonstrates near-linear scalability initially, with performance plateauing at higher parallelism levels.
This pattern is mirrored in latency metrics.
As parallelism increases, throughput improves but latency rises, indicating diminishing returns.
This tradeoff underscores the importance of careful framework optimization and configuration to achieve optimal performance on the given hardware according to the usecase.

\begin{figure}
    \centering
    \begin{subfigure}[b]{0.33\textwidth}
        \centering
        \includegraphics[width=\textwidth]{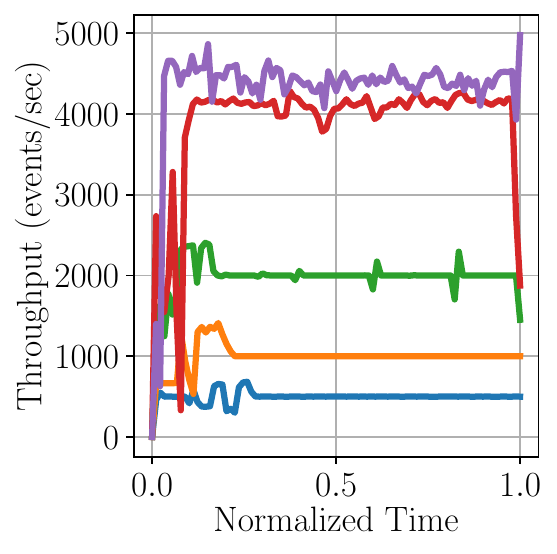}
        \caption{Runtime Vs Throughput}
        \label{fig:scaleup_res_rt_a}
    \end{subfigure}~
    \begin{subfigure}[b]{0.33\textwidth}
        \centering
        \includegraphics[width=\textwidth]{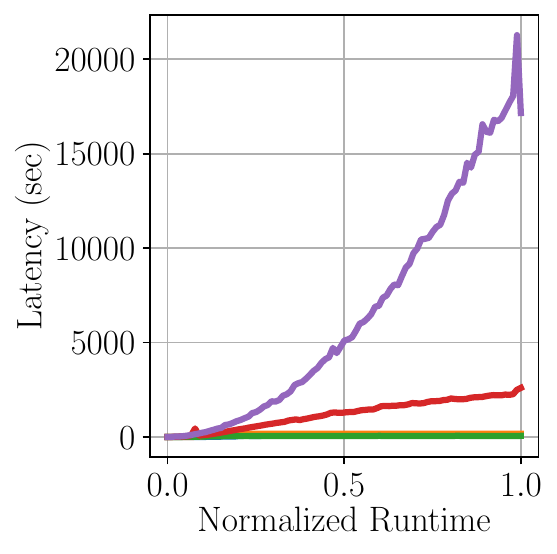}
        \caption{Runtime Vs Latency}
        \label{fig:scaleup_res_rt_b}
    \end{subfigure}~
    \begin{subfigure}[b]{0.37\textwidth}
        \centering
        \includegraphics[width=\textwidth]{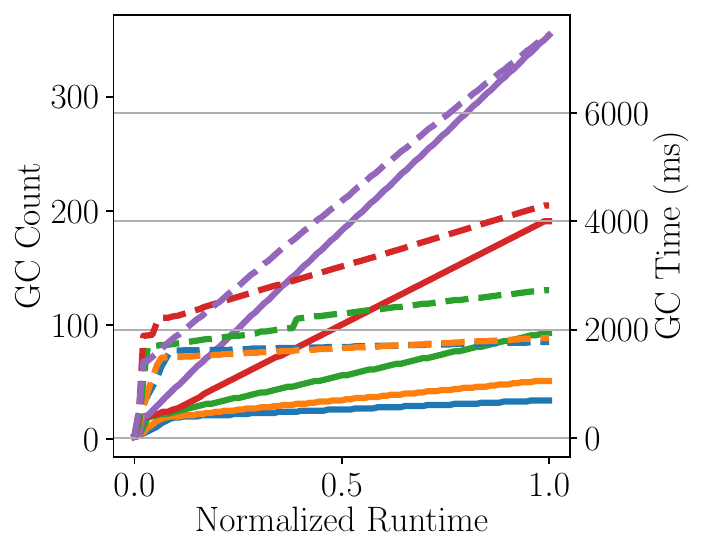}
        {\tiny - Count, - - Time}
        \caption{Runtime Vs GC (Young)}
        \label{fig:scaleup_res_rt_e}
    \end{subfigure}\\
    {Parallelism: \textbf{\color{C0} $\boldsymbol{-}$} 1, \textbf{\color{C1} $\boldsymbol{-}$} 2,\textbf{\color{C2} $\boldsymbol{-}$} 4,\textbf{\color{C3} $\boldsymbol{-}$} 8, \textbf{\color{C4} $\boldsymbol{-}$} 16}
    \caption{Metrics across normalized runtime}
    \label{fig:metrics_runtime}
\end{figure}

Figure~\ref{fig:metrics_runtime} illustrates the benchmark's performance metrics at various levels of parallelism (1, 2, 4, 8, and 16 threads, represented by different colored lines) throughout the runtime.
Figure~\ref{fig:scaleup_res_rt_a} and \ref{fig:scaleup_res_rt_b} display how throughput and latency change over normalized runtime.
It can be observed that, higher parallelism (purple line, 16 CPUs) achieves the highest throughput but also causes increasing latency compared to lower thread counts.
Figure~\ref{fig:scaleup_res_rt_e} presents garbage collection (GC) metrics, demonstrating the rise in both GC count and duration over time, indicating that higher levels of parallelism typically necessitate increased garbage collection activity.
These figures highlight the tradeoffs in the stream processing: while higher parallelism counts can improve throughput, they also introduce higher latency penalties and increased resource consumption.

\section{Conclusion and Future Works}
\label{sec:concludion}
This work introduces SProBench, a novel, modular, and highly scalable data stream processing benchmark suite.
It is designed to align with the latest advancements in stream processing frameworks and leverage the capabilities of modern HPC systems.
Effective assessment of DSP frameworks, which benefit from the vast resources and high computational capacity of modern computing systems, requires benchmarking tools capable of fully utilizing available resources and pushing both hardware and software to their limits.
As summarized in Table~\ref{tab:benchmark_comparison}, existing benchmark suites face certain challenges in execution efficiency and may introduce performance constraints.
These factors can impact the precise assessment of DSP framework capabilities.
Addressing this gap, SProBench emerges as a highly efficient and scalable solution.
SProBench demonstrates exceptional performance, with a single instance of its workload generator outperforming most existing benchmark suites.
When utilizing parallel instances, SProBench's throughput exceeds that of all other benchmark suites by more than tenfold, showcasing its remarkable efficiency and scalability.
In addition, fully configurable workloads and pipelines, native support of Slurm and popular DSP frameworks, automatic experiment workflow management, feature-rich post-processing capabilities, and an open-source software stack make SProBench stand out among the DSP benchmarking suites.

Future plans include the comprehensive integration of other widely-used DSP frameworks and conducting large-scale benchmarking on real-world clusters.
Additionally, more pre-defined pipelines and workloads will be developed to further enhance the benchmarking capabilities.

\begin{credits}
    \subsubsection{\ackname} The authors acknowledge the financial support by the Federal Ministry of Education and Research
    of Germany and by Sächsische Staatsministerium für Wissenschaft, Kultur und Tourismus in the
    programme Center of Excellence for AI-research "Center for Scalable Data Analytics and Artificial
    Intelligence Dresden/Leipzig", project identification number: ScaDS.AI.
\end{credits}

\bibliographystyle{splncs04}

\end{document}